\documentclass[epj]{webofc}%
\usepackage[varg]{txfonts}
\usepackage{amsmath}
\usepackage{amsfonts}
\usepackage{amssymb}
\usepackage{graphicx}%
\woctitle{Self Sustained Traversable Wormholes Induced by Gravity's Rainbow and Noncommutative Geometry}
\begin{document}

\title{Self Sustained Traversable Wormholes Induced by Gravity's Rainbow and
Noncommutative Geometry}

\author{Remo Garattini\inst{1,2}\fnsep\thanks{\email{remo.garattini@unibg.it}}}
\abstract{We compare the effects of Noncommutative Geometry and Gravity's Rainbow on traversable wormholes
which are sustained by their own gravitational quantum fluctuations. Fixing the geometry on a well tested model,
we find that the final result shows that the wormhole is of the Planckian size.
This means that the traversability of the wormhole is in principle, but not in practice.}

\institute{Università degli Studi di Bergamo, Facoltà di Ingegneria,\\
Viale Marconi,5 24044 Dalmine (Bergamo) ITALY\\
\and
I.N.F.N. - sezione di Milano, Milan, Italy}
\maketitle
\section{Introduction}

\label{intro}In 1988 on the American Journal of Physics, M. S. Morris and K.
S. Thorne published a paper entitled \textquotedblleft\textit{Wormholes in
spacetimes and their use for interstellar travel: A tool for teaching general
relativity}\textquotedblright\cite{MT}. Although the subject of the paper
could be regarded as an argument of Science Fiction more than of Science, its
impact on the scientific community was so amazing to open the doors to new
investigations in Astrophysics, General Relativity and Quantum Gravity. In
practice a traversable wormhole is a solution of the Einstein's Field
equations, represented by two asymptotically flat regions joined by a bridge:
roughly speaking it looks like a short-cut in space and time. To exist,
traversable wormholes must violate the null energy conditions, which means
that the matter threading the wormhole's throat has to be \textquotedblleft%
\textit{exotic}\textquotedblright. Classical matter satisfies the usual energy
conditions. Therefore, it is likely that wormholes must belong to the realm of
semiclassical or perhaps a possible quantum theory of the gravitational field.
Since a complete theory of quantum gravity has yet to come, it is important to
approach this problem semiclassically. On this ground, the Casimir energy on a
fixed background. has the correct properties to substitute the exotic matter:
indeed, it is known that, for different physical systems, Casimir energy is
negative. However, instead of studying the Casimir energy contribution of some
matter or gauge fields to the traversability of the wormholes, we propose to
use the energy of the graviton on a background of a traversable wormhole. In
this way, one can think that the quantum fluctuations of the traversable
wormholes can be used as a fuel to sustain traversability. Different contexts
can be invoked to study self sustained traversable wormholes. In this paper,
we review some aspects of self sustained traversable wormholes fixing our
attention on Noncommutative geometry and Gravity's Rainbow.

\section{Self-sustained Traversable Wormholes}

\label{sec-1}In this Section we shall consider the formalism outlined in
detail in Refs. \cite{Remo,Remo1}, where the graviton one loop contribution to
a classical energy in a wormhole background is used. The spacetime metric
representing a spherically symmetric and static wormhole is given by
\begin{equation}
ds^{2}=-e^{2\Phi(r)}\,dt^{2}+\frac{dr^{2}}{1-b(r)/r}+r^{2}\,(d\theta^{2}%
+\sin^{2}{\theta}\,d\phi^{2})\,, \label{metricwormhole}%
\end{equation}
where $\Phi(r)$ and $b(r)$ are arbitrary functions of the radial coordinate,
$r$, denoted as the redshift function, and the form function, respectively
\cite{MT}. The radial coordinate has a range that increases from a minimum
value at $r_{0}$, corresponding to the wormhole throat, to infinity. A
fundamental property of a wormhole is that a flaring out condition of the
throat, given by $(b-b^{\prime}r)/b^{2}>0$, is imposed \cite{MT,Visser}, and
at the throat $b(r_{0})=r=r_{0}$, the condition $b^{\prime}(r_{0})<1$ is
imposed to have wormhole solutions. Another condition that needs to be
satisfied is $1-b(r)/r>0$. For the wormhole to be traversable, one must demand
that there are no horizons present, which are identified as the surfaces with
$e^{2\Phi}\rightarrow0$, so that $\Phi(r)$ must be finite everywhere. The
classical energy is given by
\[
H_{\Sigma}^{(0)}=\int_{\Sigma}\,d^{3}x\,\mathcal{H}^{(0)}=-\frac{1}{16\pi
G}\int_{\Sigma}\,d^{3}x\,\sqrt{g}\,R\,,
\]
where the background field super-hamiltonian, $\mathcal{H}^{(0)}$, is
integrated on a constant time hypersurface. $R$ is the curvature scalar, and
using metric $\left(  \ref{metricwormhole}\right)  $, is given by
\[
R=-2\left(  1-\frac{b}{r}\right)  \left[  \Phi^{\prime\prime}+(\Phi^{\prime
})^{2}-\frac{b^{\prime}}{r(r-b)}-\frac{b^{\prime}r+3b-4r}{2r(r-b)}%
\,\Phi^{\prime}\right]  \,.
\]
We shall henceforth consider a constant redshift function, $\Phi^{\prime
}(r)=0$, which provides interestingly enough results, so that the curvature
scalar reduces to $R=2b^{\prime}/r^{2}$. Thus, the classical energy reduces
to
\begin{equation}
H_{\Sigma}^{(0)}=-\frac{1}{2G}\int_{r_{0}}^{\infty}\,\frac{dr\,r^{2}}%
{\sqrt{1-b(r)/r}}\,\frac{b^{\prime}(r)}{r^{2}}\,. \label{classical}%
\end{equation}
A traversable wormhole is said to be \textquotedblleft\textit{self
sustained}\textquotedblright\ if%
\begin{equation}
H_{\Sigma}^{(0)}=-E^{TT}, \label{SS}%
\end{equation}
where $E^{TT}$ is the total regularized graviton one loop energy. Basically
this is given by
\begin{equation}
E^{TT}=-\frac{1}{2}\sum_{\tau}\left[  \sqrt{E_{1}^{2}\left(  \tau\right)
}+\sqrt{E_{2}^{2}\left(  \tau\right)  }\right]  \,,
\end{equation}
where $\tau$ denotes a complete set of indices and $E_{i}^{2}\left(
\tau\right)  >0$, $i=1,2$ are the eigenvalues of the modified Lichnerowicz
operator%
\begin{equation}
\left(  \hat{\bigtriangleup}_{L\!}^{m}\!{}\;h^{\bot}\right)  _{ij}=\left(
\bigtriangleup_{L\!}\!{}\;h^{\bot}\right)  _{ij}-4R{}_{i}^{k}\!{}%
\;h_{kj}^{\bot}+\text{ }^{3}R{}\!{}\;h_{ij}^{\bot}\,,
\end{equation}
acting on traceless-transverse tensors of the perturbation and where
$\bigtriangleup_{L}$is the Lichnerowicz operator defined by%
\begin{equation}
\left(  \bigtriangleup_{L}\;h\right)  _{ij}=\bigtriangleup h_{ij}%
-2R_{ikjl}\,h^{kl}+R_{ik}\,h_{j}^{k}+R_{jk}\,h_{i}^{k},
\end{equation}
with $\bigtriangleup=-\nabla^{a}\nabla_{a}$. For the background $\left(
\ref{metricwormhole}\right)  $, one can define two r-dependent radial wave
numbers%
\begin{equation}
k_{i}^{2}\left(  r,l,\omega_{i,nl}\right)  =\omega_{i,nl}^{2}-\frac{l\left(
l+1\right)  }{r^{2}}-m_{i}^{2}\left(  r\right)  \quad i=1,2\quad, \label{kTT}%
\end{equation}
where%
\begin{equation}
\left\{
\begin{array}
[c]{c}%
m_{1}^{2}\left(  r\right)  =\frac{6}{r^{2}}\left(  1-\frac{b\left(  r\right)
}{r}\right)  +\frac{3}{2r^{2}}b^{\prime}\left(  r\right)  -\frac{3}{2r^{3}%
}b\left(  r\right) \\
\\
m_{2}^{2}\left(  r\right)  =\frac{6}{r^{2}}\left(  1-\frac{b\left(  r\right)
}{r}\right)  +\frac{1}{2r^{2}}b^{\prime}\left(  r\right)  +\frac{3}{2r^{3}%
}b\left(  r\right)
\end{array}
\right.  \label{masses}%
\end{equation}
are two r-dependent effective masses $m_{1}^{2}\left(  r\right)  $ and
$m_{2}^{2}\left(  r\right)  $. When we perform the sum over all modes,
$E^{TT}$ is usually divergent. In Refs. \cite{Remo,Remo1} a zeta
regularization and a renormalization have been adopted to handle the
divergences. In this paper, we will consider the effect of the Noncommutative
geometry and Gravity's Rainbow on the graviton to one loop.

\section{Noncommutative Geometry and Gravity's Rainbow at work on a
Traversable Wormhole Background}

One of the purposes of Eq.$\left(  \ref{SS}\right)  $ is the possible
discovery of a traversable wormhole with the determination of the shape
function. Nevertheless, another strategy can be considered if we fix the
wormhole shape to be traversable, at least in principle. One good candidate is%
\begin{subequations}
\begin{equation}
b\left(  r\right)  =r_{0}^{2}/r,\label{b(r)}%
\end{equation}
which is the prototype of the traversable wormholes\cite{MT}. Plugging the
shape function $\left(  \ref{b(r)}\right)  $ into Eq.$\left(  \ref{SS}\right)
$, we find that the left hand side becomes%
\end{subequations}
\begin{equation}
H_{\Sigma}^{(0)}=\frac{1}{2G}\int_{r_{0}}^{\infty}\,\frac{dr\,r^{2}}%
{\sqrt{1-r_{0}^{2}/r^{2}}}\,\frac{r_{0}^{2}}{r^{4}}\,,
\end{equation}
while the right hand side is divergent. To handle with divergences, we have
several possibilities. In this paper we adopt and compare two schemes: the
Noncommutative scheme and the Gravity's Rainbow procedure. Beginning with the
Noncommutative scheme, we recall that in Ref.\cite{RGPN}, we used the
distorted number of states%
\begin{equation}
dn_{i}=\frac{d^{3}\vec{x}d^{3}\vec{k}}{\left(  2\pi\right)  ^{3}}\exp\left(
-\frac{\theta}{4}\left(  \omega_{i,nl}^{2}-m_{i}^{2}\left(  r\right)  \right)
\right)  ,\quad i=1,2\label{moddn}%
\end{equation}
to compute the graviton one loop contribution to a cosmological constant. The
distortion induced by the Noncommutative space time allows the right hand side
of Eq.$\left(  \ref{SS}\right)  $ to be finite. Indeed, plugging $dn_{i}$ into
Eq.$\left(  \ref{SS}\right)  $, one finds that the self sustained equation for
the energy density becomes%
\begin{equation}
\frac{3\pi^{2}}{Gr_{0}^{2}}=\int_{0}^{+\infty}\sqrt{\left(  \omega^{2}%
+\frac{3}{r_{0}^{2}}\right)  ^{3}}e^{-\frac{\theta}{4}\left(  \omega^{2}%
+\frac{3}{r_{0}^{2}}\right)  }d\omega+\int_{1/r_{0}}^{+\infty}\sqrt{\left(
\omega^{2}-\frac{1}{r_{0}^{2}}\right)  ^{3}}e^{-\frac{\theta}{4}\left(
\omega^{2}-\frac{1}{r_{0}^{2}}\right)  }d\omega,\label{Schw1loop}%
\end{equation}
where we have used the shape function $\left(  \ref{b(r)}\right)  $ to
evaluate the effective masses $\left(  \ref{masses}\right)  $. If we define
the dimensionless variable%
\begin{equation}
x=\frac{\theta}{4r_{0}^{2}},
\end{equation}
Eq.$\left(  \ref{Schw1loop}\right)  $ leads to $\left(  G=l_{P}^{2}\right)  $%
\[
\frac{3\pi^{2}\theta}{l_{P}^{2}}=F\left(  x\right)  ,
\]
where%
\begin{equation}
F\left(  x\right)  =\left(  \left(  1-x\right)  K_{1}\left(  \frac{x}%
{2}\right)  +xK_{0}\left(  \frac{x}{2}\right)  \right)  \exp\left(  \frac
{x}{2}\right)  +3\left(  \left(  1+3x\right)  K_{1}\left(  \frac{3x}%
{2}\right)  +3xK_{0}\left(  \frac{3x}{2}\right)  \right)  \exp\left(
-\frac{3x}{2}\right)  .
\end{equation}
$F\left(  x\right)  $ has a maximum for $\bar{x}=0.24$, where%
\begin{equation}
\frac{3\pi^{2}\theta}{l_{P}^{2}}=F\left(  \bar{x}\right)  =2.20.
\end{equation}
This fixes $\theta$ to be%
\begin{equation}
\theta=\frac{2.20l_{P}^{2}}{3\pi^{2}}=7.\,\allowbreak43\times10^{-2}l_{P}%
^{2}.\label{theta}%
\end{equation}
and%
\begin{equation}
r_{0}=0.28l_{P}.
\end{equation}
As regards Gravity's Rainbow, as shown in Ref.\cite{RGFSNL}, the self
sustained equation $\left(  \ref{SS}\right)  $ becomes
\begin{equation}
\frac{b^{\prime}(r)}{2Gg_{2}\left(  E/E_{P}\right)  r^{2}}=\frac{2}{3\pi^{2}%
}\left(  I_{1}+I_{2}\right)  \,,\label{ETT}%
\end{equation}
where%
\begin{equation}
I_{1}=\int_{E^{\ast}}^{\infty}E\frac{g_{1}\left(  E/E_{P}\right)  }{g_{2}%
^{2}\left(  E/E_{P}\right)  }\frac{d}{dE}\left(  \frac{E^{2}}{g_{2}^{2}\left(
E/E_{P}\right)  }-m_{1}^{2}\left(  r\right)  \right)  ^{\frac{3}{2}%
}dE\,,\label{I1}%
\end{equation}
and%
\begin{equation}
I_{2}=\int_{E^{\ast}}^{\infty}E\frac{g_{1}\left(  E/E_{P}\right)  }{g_{2}%
^{2}\left(  E/E_{P}\right)  }\frac{d}{dE}\left(  \frac{E^{2}}{g_{2}^{2}\left(
E/E_{P}\right)  }-m_{2}^{2}\left(  r\right)  \right)  ^{\frac{3}{2}%
}dE\,.\label{I2}%
\end{equation}
Eq.$\left(  \ref{ETT}\right)  $ is finite for appropriate choices of the
Rainbow's functions $g_{1}\left(  E/E_{P}\right)  $ and $g_{2}\left(
E/E_{P}\right)  $. Fixing the shape function as in Eq.$\left(  \ref{b(r)}%
\right)  $ and assuming $g_{2}\left(  E/E_{P}\right)  =1$, to avoid Planckian
distortions in the classical term, we find%
\begin{equation}
I_{1}=3\int_{\sqrt{3/r_{t}^{2}}}^{\infty}\exp\left(  -\alpha E^{2}/E_{P}%
^{2}\right)  E^{2}\sqrt{E^{2}-\frac{3}{r_{t}^{2}}}dE\label{I1a}%
\end{equation}
and%
\begin{equation}
I_{2}=3\int_{0}^{\infty}\exp\left(  -\alpha E^{2}/E_{P}^{2}\right)  E^{2}%
\sqrt{E^{2}+\frac{1}{r_{t}^{2}}}dE\,,\label{I1b}%
\end{equation}
where we have also fixed $g_{1}(E/E_{P})=\exp\left(  -\alpha E^{2}/E_{P}%
^{2}\right)  $ with $\alpha$ variable.

Now, in order to have only one solution with variables $\alpha$ and $r_{t}$,
we demand that%
\begin{equation}
\frac{d}{dr_{t}}\left[  -\frac{1}{2G}\,\frac{1}{r_{t}^{2}}\right]  \,=\frac
{d}{dr_{t}}\left[  \frac{2}{3\pi^{2}}\left(  I_{1}+I_{2}\right)  \right]  \,,
\end{equation}
which takes the following form after the integration $\left(  G^{-1}=E_{P}%
^{2}\right)  $%
\begin{equation}
1=\frac{1}{2\pi^{2}x^{2}}f\left(  \alpha,x\right)  \,,\label{SSEq1}%
\end{equation}
where $x=r_{0}E_{P}$ and where
\begin{equation}
f\left(  \alpha,x\right)  =\exp\left(  \frac{\alpha}{2x^{2}}\right)
K_{0}\left(  \frac{\alpha}{2x^{2}}\right)  -\exp\left(  \frac{\alpha}{2x^{2}%
}\right)  K_{1}\left(  \frac{\alpha}{2x^{2}}\right)  +9\exp\left(
-\frac{3\alpha}{2x^{2}}\right)  K_{0}\left(  \frac{3\alpha}{2x^{2}}\right)
+\exp\left(  -\frac{3\alpha}{2x^{2}}\right)  K_{1}\left(  \frac{3\alpha
}{2x^{2}}\right)  .
\end{equation}
$K_{0}(x)$ and $K_{1}(x)$ are the modified Bessel function of order 0 and 1,
respectively. Even in this case, to have one and only one solution, we demand
that the expression in the right hand side of Eq. $(\ref{SSEq1})$ has a
stationary point with respect to $x$ which coincides with the constant value
$1$. For a generic but small $\alpha$, we can expand in powers of $\alpha$ to
find%
\begin{equation}
0=\frac{d}{dx}\left[  \frac{1}{2\pi^{2}x^{2}}f\left(  \alpha,x\right)
\right]  \simeq\frac{20-10\ln\left(  4x^{2}/\alpha\right)  +10\gamma_{E}%
+9\ln3}{\pi^{2}x^{2}}+O\left(  \alpha\right)  \,,
\end{equation}
which has a root at%
\begin{equation}
\bar{x}=r_{t}E_{P}=2.973786871\sqrt{\alpha}\,.\label{xmin}%
\end{equation}
Substituting $\bar{x}$ into Eq. $(\ref{SSEq1})$, we find%
\begin{equation}
1=\frac{0.2423530631}{\alpha}\,,
\end{equation}
fixing therefore $\alpha\simeq0.242$. It is interesting to note that this
value is very close to the value $\alpha=1/4$ used in Ref.\cite{RGGM} inspired
by Noncommutative analysisof Ref.\cite{RGPN}\textbf{.} As in Refs.
\cite{Remo1,RGFSNLPLB}, it is rather important to emphasize a shortcoming in
the analysis carried in this section, mainly due to the technical difficulties
encountered. Note that we have considered a variational approach which imposes
a local analysis to the problem, namely, we have restricted our attention to
the behavior of the metric function $b(r)$ at the wormhole throat, $r_{0}$.
Despite the fact that the behavior is unknown far from the throat, due to the
high curvature effects at or near $r_{0}$, the analysis carried out in this
section should extend to the immediate neighborhood of the wormhole throat.
Nevertheless it is interesting to observe that in Ref.\cite{Remo} the greatest
value of the wormhole throat was fixed at $r_{0}\simeq1.16/E_{P}$ using a
regularization-renormalization scheme. From Eq.$\left(  \ref{xmin}\right)  $,
one immediately extracts $r_{0}\simeq1.46/E_{P}$ which is slightly larger. We
have to remark that in the Noncommutative case we have only one parameter to
be fixed: $\theta=7.\,\allowbreak43\times10^{-2}l_{P}^{2}$. On the other hand,
in Gravity's Rainbow, we have more flexibility, because of the unknown
functions $g_{1}\left(  E/E_{P}\right)  $ and $g_{2}\left(  E/E_{P}\right)  $.
Therefore we conclude that Gravity's Rainbow offers a wider variety of
examples which deserve to be explored.

\end{document}